\documentstyle[prl,aps,multicol,subfigure,psfig,subeqnar]{revtex}

\newcommand{\beq}{\begin{equation}}
\newcommand{\eeq}{\end{equation}}
\newcommand{\ba}{\begin{array}}
\newcommand{\ea}{\end{array}}
\newcommand{\bea}{\begin{eqnarray}}
\newcommand{\eea}{\end{eqnarray}}
\newcommand{\bc}{\begin{center}}
\newcommand{\ec}{\end{center}}
\newcommand{\bt}{\begin{table}}
\newcommand{\et}{\end{table}}

\newcommand{\la}[1]{\label{#1}}
\newcommand{\p}{\partial}

\newcommand{\no}{\noindent}

\newcommand{\mbf}[1]{\mbox{\boldmath {$#1$}}}

\newcommand{\rf}[1]{(\ref{#1})}
\newcommand{\beqno}{\begin{displaymath}}
\newcommand{\eeqno}{\end{displaymath}}

\newcommand{\been}{\begin{enumerate}}
\newcommand{\een}{\end{enumerate}}

\newlength{\myheight}
\newlength{\mylength}

\newcounter{saveeqn}

\newtheorem{example}{Example}

\begin{document}

\title{Singular instability of exact stationary solutions of
   the nonlocal Gross-Pitaevskii equation}
%Nonlocal mean-field theory in $N$-body quantum
%mechanics: applications to Bose-Einstein condensates in standing light waves}
%
\author{Bernard Deconinck$^{1}$, and J. Nathan Kutz$^{2}$\\}
\address{$^{1}$Department of Mathematics, Colorado State University, 
        Fort Collins, CO 80523, USA\\}
\address{$^{2}$Department of Applied Mathematics, University of 
         Washington, Seattle, WA 98195-2420, USA\\}
\maketitle

\date{\today}

\begin{abstract}
  In this paper we show numerically that for nonlinear Schr\"odinger
  type systems the presence of nonlocal
  perturbations can lead to a beyond-all-orders instability of stable
  solutions of the local equation.  For the specific case of the nonlocal
  one-dimensional Gross-Pitaevskii equation with an external standing light
  wave potential, we construct exact stationary solutions for an arbitrary
  interaction kernel.  As the nonlocal and local equations approach each other
  (by letting an appropriate small parameter $\epsilon\rightarrow 0$), we
  compare the dynamics of the respective solutions.  By considering the time
  of onset of instability, the singular nature of the inclusion of nonlocality
  is demonstrated, independent of the form of the interaction kernel.
\end{abstract}

% PACS numbers??

\begin{multicols}{2}
  
  In almost all applications where the nonlinear Schr\"odinger (NLS) equation
  is relevant, it arises as a simplified model of a nonlocal description.
  This is true in water waves, plasma physics, 
  nonlinear optics, and
  Bose-Einstein condensates (BECs).  For water waves~\cite{water} 
  and plasmas~\cite{plasma}, the
  nonlocality occurs due to a Fourier transformation of nonlinear equations.
  In nonlinear optics~\cite{optics}, the nonlocality is due to the spatial dependence of
  the susceptibility tensor.  Locality in each of these cases is obtained as a
  quasi-monochromatic approximation of the nonlocal model.  In plasma physics
  the nonlocal effect is known as Landau damping.  The use of mean-field
  theory results in nonlocality for BECs~\cite{Baym,Gross,Pitaevskii}, 
  which reduces to the
  Gross-Pitaevskii (GP) equation (the NLS equation with external potential)
  when assuming a hard pairwise interaction potential.  As such, the
  consideration of nonlocal perturbations of solutions of the NLS equation is
  important.  Such perturbations result in the NLS equation with the cubic
  nonlinearity replaced by a nonlocal, nonlinear term.  The nonlocal,
  nonlinear term is a convolution of the modulus squared of the
  solution with an interaction kernel, prescribed by the physical problem.  In
  this paper our objective is to numerically examine the stability of
  solutions of this nonlocal NLS equation, and discuss how it differs from
  the local description.
  
  To consider a specific nonlocal model, we examine mean-field theory of
  many-particle quantum mechanics with the particular application of BECs
  trapped in a standing light wave.  The classical derivation given here is
  included to illustrate how the local and nonlocal models are related.  The
  inherent complexity of the dynamics of $N$ pairwise interacting particles in
  quantum mechanics often leads to the consideration of such simplified
  mean-field descriptions.  These descriptions are a blend of symmetry
  restrictions on the particle wave function \cite{Baym} and functional form
  assumptions on the interaction potential \cite{Baym,Gross,Pitaevskii}.
  Here we do not impose any assumptions on the pairwise interaction potential.
  
The dynamics of $N$ identical pairwise interacting quantum particles is
governed by the time-dependent, $N$-body Schr\"odinger equation
\begin{equation}
  i \hbar \frac{\partial \Psi}{\partial t} \!\!=\!\! -\frac{\hbar^2}{2 m} 
  \Delta^N \Psi \!+\! 
    \sum_{i=1}^{N} \!W({\mbf x}_i\!-\!\mbf{x}_j) \Psi \!+\! \sum_{i=1}^N \!V(\mbf{x_i})
    \Psi,
\label{eq:schro}
\end{equation}
where $\mbf{x_i}=(x_{i1},x_{i2},x_{i3})$,
%$i=1,\ldots, N$, 
$\Psi=\Psi(\mbf{x}_1,\mbf{x}_2,\mbf{x}_3,...,\mbf{x}_N,t)$ is the  wave
function of the $N$-particle system, $\Delta^N =\left(\nabla^N \right)^2 
=\sum_{i=1}^N \left(
\p^2_{x_{i1}}+\p^2_{x_{i2}}+\p_{x_{i3}}^2\right)$ is the kinetic energy or Laplacian
operator for $N$-particles,  $W(\mbf{x_i}-\mbf{x_j})$ is the 
symmetric interaction potential
between the $i$-th and $j$-th particle, and $V(\mbf{x}_i)$ is 
an external potential acting on the $i$-th particle. Also, $\hbar$ is Planck's
constant divided by $2\pi$ and $m$ is the mass of the particles under
consideration. 

One way to arrive at a mean-field description is by using the Lagrangian
reduction technique~\cite{Whittham}, which exploits the Hamiltonian structure of
Eq.~\rf{eq:schro}.  The Lagrangian of Eq.~\rf{eq:schro} is given
by~\cite{Baym}
\bea\nonumber
  L&=& \int_{-\infty}^{\infty} \left\{ i \frac{\hbar}{2}
\left( \Psi \frac{\partial \Psi^*}{\partial t} - 
       \Psi^* \frac{\partial \Psi}{\partial t} \right) +
\frac{\hbar^2}{2m}\left| \nabla^N \Psi \right|^2 \right. \\
   && \left. -\sum_{i=1}^{N} \left(W(\mbf{x}_i-\mbf{x}_j)+V(\mbf{x}_i)\right) \left| \Psi \right|^2 \right\}
d \mbf{x}_1\cdots d \mbf{x}_N
\label{eq:lagrangian}
\eea
The Hartree-Fock approximation (as used in \cite{Baym}) for bosonic particles
uses the separated wave function ansatz
\begin{equation}
  \Psi=\psi_1(\mbf{x}_1,t)\psi_2(\mbf{x}_2,t)\cdots\psi_N(\mbf{x}_N,t) 
\label{eq:separate}
\end{equation}
where each one-particle wave function $\psi(\mbf{x}_i)$ is assumed to be
normalized so that $\langle \psi(\mbf{x}_i) | \psi(\mbf{x}_i) \rangle^2=1$.
Since identical particles are being considered,
\begin{equation}
  \psi_1=\psi_2=\ldots=\psi_N=\psi,
\label{eq:symmetry}
\end{equation}
enforcing total symmetry of the wavefunction.  Note that for the case of BECs,
assumption \rf{eq:separate} is only approximate if the temperature is not
identically zero.

Integrating Eq.~(\ref{eq:lagrangian}) using (\ref{eq:separate}) and (\ref{eq:symmetry}) 
and taking the variational derivative with respect to $\psi(\mbf{x}_i)$ results in the
Euler-Lagrange equation~\cite{Whittham}
\bea\nonumber
   &\!i\hbar\!& \frac{\partial \psi(\mbf{x},t)}{\partial t}=
-\frac{\hbar^2}{2m} \Delta \psi(\mbf{x},t)
       + V(\mbf{x}) \psi(\mbf{x},t)\\
       &&~+ (N-1)\psi(\mbf{x},t)\int_{-\infty}^\infty W(\mbf{x}-\mbf{y}) 
  |\psi(\mbf{y},t)|^2 d\mbf{y}.
\label{eq:mean}
\eea
Here, $\mbf{x}=\mbf{x}_i$, and $\Delta$ is the one-particle Laplacian in three
dimensions.  The Euler-Lagrange equation \rf{eq:mean} is identical
for all $\psi(\mbf{x}_i,t)$.  Equation \rf{eq:mean} describes the nonlinear,
nonlocal, mean-field dynamics of the wave function $\psi(\mbf{x},t)$ under
the standard assumptions~\rf{eq:separate} and \rf{eq:symmetry} of
Hartree-Fock theory \cite{Baym}. The coefficient of $\psi(x,t)$ in the last
term in Eq.~\rf{eq:mean} represents the effective potential acting on
$\psi(x,t)$ due to the presence of the other particles.

At this point, it is common to make an assumption on the functional form of
the interaction potential $W(\mbf{x}-\mbf{y})$. This is done to render
Eq.~\rf{eq:mean} analytically and numerically tractable. Although the
qualitative features of this functional form may be available, for instance
from experiment, its quantitative details are rarely known. One convenient
assumption in the case of short-range potential interactions is
$W(\mbf{x}-\mbf{y})=\kappa \delta(\mbf{x}-\mbf{y})$ where $\delta$ is the
Dirac delta function. This leads to the Gross-Pitaevskii
\cite{Gross,Pitaevskii} mean-field description:
\begin{equation}
    i \hbar \frac{\partial \psi}{\partial t} = -\frac{\hbar^2}{2m} 
       \Delta \psi
       + \beta |\psi|^2 \psi
       + V(\mbf{x}) \psi,
\label{eq:gp}
\end{equation}
where $\beta=(N-1)\kappa$ reflects whether the interaction is repulsive
($\beta>0$) or attractive ($\beta<0$).  This assumption on the interaction
potential $W(\mbf{x}-\mbf{y})$ is difficult to physically justify.
Nevertheless, Lieb and Seiringer~\cite{Lieb} show that Eq.~(\ref{eq:gp})
is the correct asymptotic description in the dilute-gas limit. 
In this limit, Eqs.~\rf{eq:mean} and \rf{eq:gp} are asymptotically equivalent.
Thus the nonlocal Eq.~\rf{eq:gp} can be interpreted as a perturbation to 
the local Eq.~\rf{eq:mean}.  Note that the results of ~\cite{Lieb} do not
have implications for the asymptotic equivalence of the stability of solutions.

% REDUN
%The considerations leading to Eqs.~\rf{eq:mean} and
%\rf{eq:gp} are well known. 
%They are included here since one of the aims of
%this letter is to emphasize the differences between these two mean-field
%theories.  
%
%This construction is best illustrated with an example, which we
%take from Bose-Einstein condensation.  Bose-Einstein condensates (BECs) are
%often described by the GP equation \rf{eq:gp} \cite{bunch}. The limited
%validity of this equation leads us to reconsider Eq.~\rf{eq:mean}, without
%imposing conditions on $W(\mbf{x}-\mbf{y})$. 

Since their first successful demonstrations in
1995~\cite{Cornell,Ketterle}, continuous progress is being made in
trapping, controlling, and manipulating Bose-Einstein condensates in a variety
of experimental configurations~\cite{becbook}.  Although many experiments rely
solely on harmonic confinement to trap the condensate, we consider the
situation of an external standing-light wave potential within a confining
potential~\cite{Kasevich,prl,us}.  This standing-light wave pattern is
generated by the interference of two quasi-monochromatic lasers in a
quasi-one-dimensional configuration.  The quasi-one-dimensional regime holds
when the transverse dimensions of the condensate are on the order of its
healing length and its longitudinal dimension is much longer than the
transverse ones. The rescaled governing mean-field evolution~\rf{eq:mean} in
the quasi-one-dimensional regime is given by
\begin{equation}
   i \frac{\partial \psi}{\partial t}\!=\!
  -\frac{1}{2} \frac{\partial^2 \psi}{\partial x^2}
       \!+\! \alpha \psi \!\! \int_{-\infty}^\infty \!\!\!\!\!\!\! R(x-y) 
  |\psi(y,t)|^2 dy \!+\! V(x) \psi.
\label{eq:1d}
\end{equation}
Here $\alpha=\pm 1$ is the sign of the interaction potential $W(x-y)$ at close
range. Thus, $\alpha$ determines whether the close-range interaction is
repulsive ($\alpha=1$), or attractive ($\alpha=-1$).  Hence,
$\alpha={\rm sign}(a)$, where $a$ is the $s$-wave scattering length of the
atomic species. Depending on the species, $a$ is either positive or negative, 
so that both signs of $\alpha={\rm sign}(a)$ are relevant for BEC applications. 
With these defintions, $R(x-y)$ is the
rescaled interaction potential, which is positive at close range and
is normalized to unity $\int_{-\infty}^{\infty}R(z) dz=1$.   
This normalization condition is equivalent to a rescaling of variables.
%The
%dimensionless variables in (\ref{eq:mean}) are related to the physical
%variables by $t\rightarrow (4\pi \hbar |a| N/m\Omega)t$, $x\rightarrow
%(\Omega/4\pi |a| N)^{1/2} x$, $\psi\rightarrow (\hbar\Omega)^{-1/2}\psi$, and
%$V(x)\rightarrow (m\Omega/4\pi |a| N) V(x)$.  Here $a$ is the $s$-wave
%scattering length for collisions between atoms.  

The external potential which models the standing light wave is given by
\cite{becbook,us}
\begin{equation}
  V(x)=V_0 \sin^2(k x)
\label{eq:pot}
\end{equation}
where $k$ is the wavelength of the periodic potential. 

The nonlocal, nonlinear equation~\rf{eq:1d} with the periodic potential
\rf{eq:pot} admits a one-parameter family of exact solutions.  These solutions
are found using an amplitude-phase decomposition
\begin{equation}
  \psi(x,t)=r(x) \exp \left[ i \theta(x)-i\omega t \right] \, .
  \label{eq:ansatz}
\end{equation}
Then 
\begin{equation}
  r(x)^2=A \sin^2 (kx) + B
\end{equation}
where $B$ is a free parameter and 
\begin{subeqnarray}\la{eq:first one}
  && A(k)= \frac{-V_0}{\alpha \beta(k)}, \\
  && \tan(\theta(x))=\sqrt{1-\frac{V_0}{\alpha B \beta (k)}}~\tan(k x), \\
  && \omega(k)= \frac{V_0+k^2}{2}+\alpha B -\frac{V_0 }{2 \beta(k)}, \\
  && \beta(k)=\int_{-\infty}^\infty R(z) \,\cos(2kz)\, dz = \hat{R} (2k),
\end{subeqnarray}
where $\hat{R}(2k)$ is the Fourier transform of $R(z)$ evaluated at $2k$.
Equations~\rf{eq:first one} can be verified by direct substitution, using the
addition formula for $\cos(2(y-x)+2x)$ and the fact that $R(x-y)$ is even.
This is the only essential mathematical assumption made on the interaction
potential $R(x-y)$ in obtaining this family of exact solutions, including the
normalization condition.  The solutions to GP theory \rf{eq:gp} found in
\cite{us} are easily recovered by letting $R(z)=\delta(z)$, $i.e.$, $\beta=1$.
%Note that the family of solutions
%determined by \rf{first one} contains as a special case two solutions with
%trivial phase, $i.e.$, solutions for which $\theta(x)=0$. These solutions are
%obtained by choosing $B=0$ or $B=V_0/(\alpha \beta)$. 

From Eq.~(\ref{eq:first one}d) and $\beta=V_0/\alpha A(k)$,
it follows that measuring the amplitude $A(k)$ of the oscillations
for varying external potential wavelengths $k$ leads to the construction
of the Fourier transform of the interaction potential $\hat{R} (2k)$.
Thus by inversion 
\begin{equation}
  R(z)=\frac{V_0}{2\pi\alpha} \int_{-\infty}^{\infty} 
   \frac{\cos(2kx)}{A(k)}dk \, .
\end{equation}
In principle, this gives a method to determine the pairwise particle
interaction potential experimentally.

Now, we investigate numerically the stability of solutions of both the local
\rf{eq:gp} and nonlocal \rf{eq:1d} equations without applying any additional
perturbations.  The computational procedure used is a 4th order Runge-Kutta
method in time and filtered pseudo-spectral method in space.  Our objective is
to explore the question of {\em asymptotic equivalence of stability} (AES): as
the nonlocal equation \rf{eq:1d} approaches the local equation \rf{eq:gp} does
the dynamics of the solution \rf{eq:ansatz} of the nonlocal equation converge
to the dynamics of the solution (\ref{eq:ansatz}, with $\beta=1$) of the local
equation?

\begin{figure}[t]
\centerline{\psfig{figure=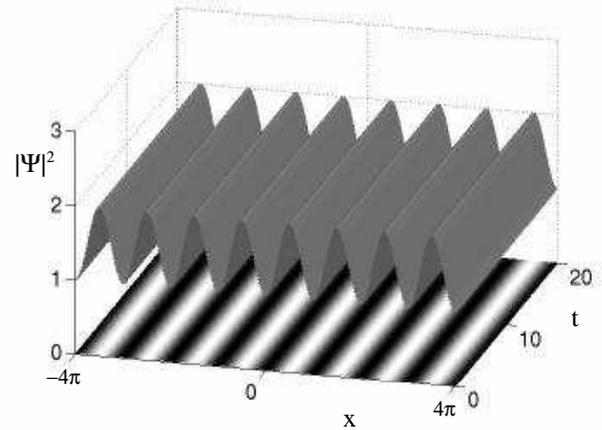,width=83mm}}
\begin{center}
\begin{minipage}{83mm}
\caption{Stable evolution of the stationary solution \rf{eq:ansatz} with
  $\alpha=1$ (repulsive), $\beta=1$ (local), $k=1$, $V_0=-1$, 
  and $B=1$.  \label{fig:stable}}
\end{minipage}
\end{center}
\end{figure}

For a sufficiently high offset value $B$, the local solution is
stable~\cite{prl,us} as illustrated in Fig.~\ref{fig:stable}.  In this
case, the time of onset of instability $t^*=\infty$, even under the influence
of large perturbations~\cite{prl,us}.  

\begin{figure}[t]
\centerline{\psfig{figure=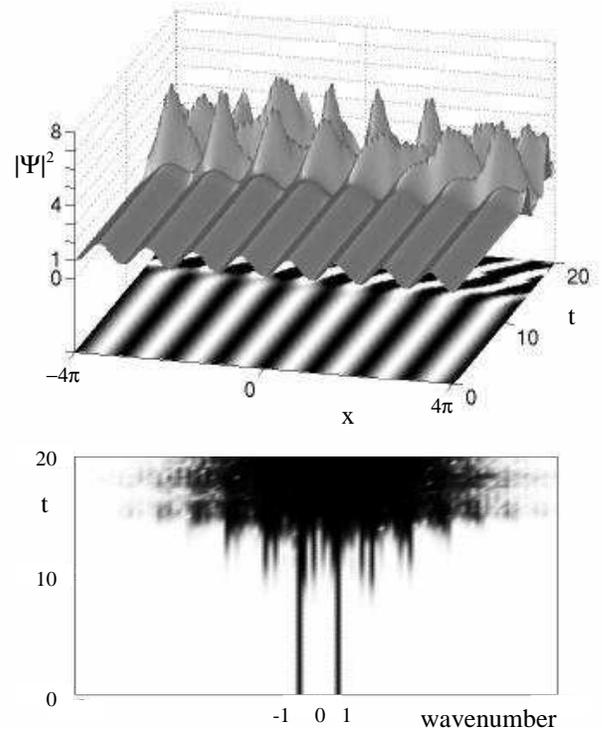,width=83mm}}
\begin{center}
\begin{minipage}{83mm}
\caption{Unstable evolution of the stationary solution \ref{eq:ansatz} with
 $\alpha=1$ (repulsive), $k=1$, $V_0=-1$, $B=1$, and $\epsilon=0.01$ in 
 Eq.~\rf{eq:gauss}.  The bottom figure is a density plot of the evolution of the
 Fourier modes.   
 \label{fig:unstable}}
\end{minipage}
\end{center}
\end{figure}

\begin{figure}[t]
\centerline{\psfig{figure=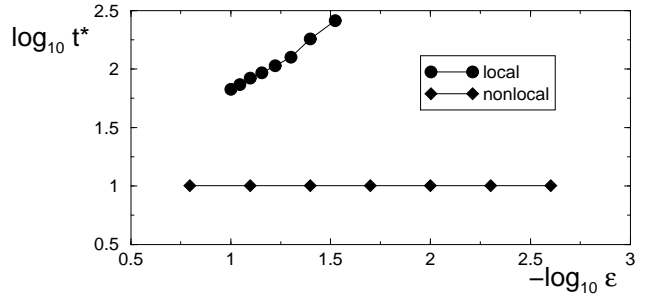,width=83mm}}
\begin{center}
\begin{minipage}{83mm}
\caption{Time of onset of instability $t^*$ as a function of $\epsilon$.
   Note that $- \log_{10} \epsilon$ gives the position of the most significant
   digit in $\epsilon$. \label{fig:time}}
\end{minipage}
\end{center}
\end{figure}

To investigate the nonlocal behavior, a choice must be made for
the interaction potential $R(z)$.  A reasonable first choice is
a Gaussian profile 
\begin{equation}
  R(z)= \frac{1}{\sqrt{2\pi}\epsilon} e^{-z^2/2\epsilon^2}
\label{eq:gauss}
\end{equation}
where as $\epsilon\rightarrow 0$, $R(z)\rightarrow \delta (z)$ and the
solution \rf{eq:ansatz} approaches the local limit. Thus examining AES 
amounts to examining $t^*$ as a function of
$\epsilon$.  The unstable dynamics and its spectral evolution are illustrated
in Fig.~\ref{fig:unstable} for $\epsilon=0.01$.  For $\epsilon\in
[0.0025,0.16]$, numerically $t^*\approx 10.1$, independent of $\epsilon$ (see
Fig.~\ref{fig:time}).
This result contradicts the expectation that $t^*\rightarrow\infty$ as
$\epsilon\rightarrow 0$ and suggests the presence of a beyond-all-orders
phenomenon.  The Fourier spectrum in Fig.~\ref{fig:unstable} provides the
primary diagnostic for studying the instability and its convergence.  Not only
does it provide the value of $t^*$, it also illustrates that the spectral
bandwidth, which is the support of the unstable modes, is independent of
$\epsilon$ for $\epsilon\in [0.0025,0.16]$.  Hence, AES is not obtained,
even in a convergence-in-measure sense.  These results raise the following
questions:  is lack of AES due to the choice of $R(z)$?  Is AES
a consequence of nonlocality or can it occur for local generalizations
of NLS?

The choice of $R(z)$ is addressed by choosing different forms of the
interaction potential, which are even and decaying at infinity.  Three
additional $R(z)$ choices were considered: $R(z)=\epsilon/(\pi(z^2 +
\epsilon^2))$ with $t^*=10.2$, $R(z)=1/(2\epsilon)$ for $x\in[-\epsilon,\epsilon]$
and $R(z)=0$ otherwise with $t^*=10.1$, and $R(z)=1/(2\epsilon)
\exp(-|z|/\epsilon)$ with $t^*=10.4$.  Thus the lack of AES appears to be a
universal feature that is independent of the interaction potential.

%%%%%%%%% figure 1 %%%%%%%%%%%%%%%%%%%%%%%%
%
%
%%%%%%%%%%%%%%%%%%%%%%%%%%%%%%%%%%%%%%%%%%

%%%%%%%%% figure 1 %%%%%%%%%%%%%%%%%%%%%%%%
%
%
%%%%%%%%%%%%%%%%%%%%%%%%%%%%%%%%%%%%%%%%%%

\begin{figure}[t]
\centerline{\psfig{figure=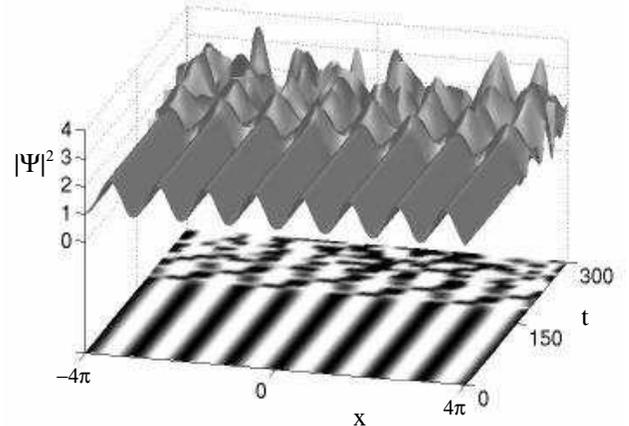,width=83mm}}
\begin{center}
\begin{minipage}{83mm}
\caption{Unstable evolution of the stationary solution \ref{eq:ansatz} with
 $\alpha=1$ (repulsive), $k=1$, $V_0=-1$, $B=1$, and $\epsilon=0.02$ in 
 Eq.~\rf{eq:bang}.  \label{fig:bang}}
\end{minipage}
\end{center}
\end{figure}

A local generalization of the NLS equation \rf{eq:1d} is
\begin{equation}
   i \frac{\partial \psi}{\partial t}\!=\!
  -\frac{1}{2} \frac{\partial^2 \psi}{\partial x^2}
       \!+\! \alpha \psi \left( |\psi|^2 +  
    \epsilon \frac{\partial^2 |\psi|^2}{\partial x^2}  \right) \!+\! V(x) \psi.
\label{eq:bang}
\end{equation}
where $\epsilon=\int_{-\infty}^{\infty} z^2 R(z)dz$.
This equation is derived by a change of variables $y=x-z$ followed by a Taylor
expansion of $|\psi|^2$ about $z=0$ in \rf{eq:1d}.  Finally assuming
$R(z)=\delta(z)$ results in \rf{eq:bang}.  Such an equation has been
considered in the multidimensional, attractive ($\alpha=-1$) case as a
successful means to arrest collapse and blow-up of
solutions~\cite{Bang,Turitsyn}.  The effect of this additional term on the
stability of solutions was never addressed.  A family of exact solutions for
this case is given by \rf{eq:ansatz} with $\beta=1-4\epsilon k^2$.  The dynamics
is found to be unstable as illustrated in Fig.~\ref{fig:bang} for
$\epsilon=0.02$.  More importantly, $\lim_{\epsilon\rightarrow 0} t^* = \infty$ as
shown in Fig.~\ref{fig:time}.  In contrast to the nonlocal Eq.~\rf{eq:1d}, the
presence of the local perturbation in Eq.~\rf{eq:bang} does not destroy AES.
This suggests that nonlocality is responsible for the beyond-all-orders
failure of AES.

Nonlocality for NLS-type equations is a generic feature arising
in physical systems.  Although nonlocality has previously been used to prevent
the non-physical features of collapse and blow-up, its effect on the stability
of solutions appear detrimental.  We have demonstrated this beyond-all-orders
singularity arising from nonlocal perturbations.  The specific model
considered is the Gross-Pitaevskii equation with a standing-light wave
potential for which an exact family of stationary solutions is constructed.
{\em Asymptotic equivalence of stability} (AES) for the nonlocal equation
\rf{eq:1d} is not achieved since the nonlocal equation dynamics does not
approach the local equation dynamics as $\epsilon\rightarrow 0$.  This is a
truly nonlocal and universal phenomenon as illustrated by the study of a
local correction model and the consideration of different interaction kernels.
The instability discussed is similar to a parametric instability. It differs
from it in that it is driven by the change of a function, not one
parameter. Further, the equations that are being compared are asymptotically
equivalent, as are their solutions. 
There is a caveat to our conclusions:  in a physical setting, many
additional effects are present which are excluded by the NLS-type model.
The success of such models can be accounted for by the presence of these
effects which may counteract the instability mechanism found here.

%In experiments, only stable solutions are expected to be observable.  Likewise,
%unstable solutions whose onset of instability occurs on a timescale beyond the
%lifetime of the experiment may be observed.  

%The analytical results are accompanied
%by direct computations on the nonlinear governing Eqs.~(\ref{eqn:NLS}) and
%(\ref{eqn:potential}).   For all computational simulations, twelve 
%spatial periods are used.  However, to better illustrate the dynamics,
%typically four spatial periods are plotted.  Moreover, all computations
%are performed with white noise included in the initial data. 

We wish to acknowledge useful discussions with Ole Bang and Harvey Segur. 
This work was supported by National Science Foundation Grants
DMS-0071568 (BD) and DMS-0092682 (JNK).

%%%%%%%%% figure 1 %%%%%%%%%%%%%%%%%%%%%%%%
%
%
%%%%%%%%%%%%%%%%%%%%%%%%%%%%%%%%%%%%%%%%%%

%%%%%%%%% figure 1 %%%%%%%%%%%%%%%%%%%%%%%%
%
%
%%%%%%%%%%%%%%%%%%%%%%%%%%%%%%%%%%%%%%%%%%

\end{multicols}
\end{document}